\documentclass[12pt]{article}
\usepackage{pic03}
\usepackage{url}
\usepackage{graphicx}
\usepackage{capt-of}

\def\Journal#1#2#3#4{{#1} {\bf #2}, #3 (#4)}

\def\PLB{{\em Phys. Lett.}  B}
\def\PRL{\em Phys. Rev. Lett.}
\def\PRD{{\em Phys. Rev.} D}

\newcommand{\ket}{\rangle}

\newcommand{\subrm}[1]{\mbox{\tiny \rm #1}}

\newcommand{\Br}{\rm Br}

\newcommand{\Reepe}{\rm{Re(}\epsilon'/\epsilon \rm{)}}

\newcommand{\reeta}{\rm{Re(}\eta_{000}\rm{)}}
\newcommand{\imeta}{\rm{Im(}\eta_{000}\rm{)}}

\newcommand{\pid}{\pi^0_{\subrm{Dalitz}}}
\newcommand{\Kbar}{\overline{K^0}}
\newcommand{\KKbar}{K^0 \overline{K^0}}
\newcommand{\kl}{K_L}
\newcommand{\ks}{K_S}
\newcommand{\kls}{K_{L,S}}

\newcommand{\kspiee}{\ks \to \pi^0 e^+ e^-}
\newcommand{\klpiee}{\kl \to \pi^0 e^+ e^-}

\newcommand{\kspipipi}{\ks \to 3 \pi^0}
\newcommand{\klpipipi}{\kl \to 3  \pi^0}

\newcommand{\klpipi}{\kl \to \pi^+ \pi^-}
\newcommand{\kspipi}{\ks \to \pi^+ \pi^-}

\newcommand{\klpizpiz}{\kl \to \pi^0 \pi^0}
\newcommand{\kspizpiz}{\ks \to \pi^0 \pi^0}

\newcommand{\ksgg}{\ks \to \gamma \gamma}

\newcommand{\klsgg}{\kls \to \gamma \gamma}

\newcommand{\kspigg}{\ks \to \pi^0 \gamma \gamma}
\newcommand{\klpigg}{\kl \to \pi^0 \gamma \gamma}

\newcommand{\kleeee}{\kl \to e^+ e^- e^+ e^-}

\newcommand{\klmumuee}{\kl \to \mu^+ \mu^- e^+ e^-}

\newcommand{\klmumu}{\kl \to \mu^+ \mu^-}
\newcommand{\kleeg}{\kl \to e^+ e^- \gamma}
\newcommand{\kleegg}{\kl \to e^+ e^- \gamma \gamma}

\newcommand{\kppienu}{K^+ \to \pi^0 e^+ \nu}

\newcommand{\kppipie}{K^+ \to \pi^+ \pi^- e^+ \nu}

\newcommand{\etazzz}{\eta_{000}}

\newcommand{\bdm}{\begin{displaymath}}
\newcommand{\edm}{\end{displaymath}}
\newcommand{\be}{\begin{equation}}
\newcommand{\ee}{\end{equation}}

\begin{document}

\title{\bf NEW RESULTS IN KAON PHYSICS FROM NA48 AND KTEV}
\author{ Rainer Wanke \\ {\em Institut f\"ur Physik, Universit\"at Mainz, D-55099 Mainz, Germany}}
\maketitle

%
%
\begin{figure}[h]
\begin{center}
%
%
%
%
\vspace{4.5cm}
\end{center}
\end{figure}

\baselineskip=14.5pt
\begin{abstract}
In the recent year many new results from kaon decays have been
reported from both the NA48 experiment at CERN and the KTeV experiment at Fermilab.
Both experiments have new and improved measurements of the parameter
$\Reepe$ of direct CP violation in the neutral kaon system.
Also several new results on rare and very rare $\kl$ decays were reported.
In addition, the NA48 collaboration 
has performed special high-intensity $\ks$ run periods, with the first results
now being available.
\end{abstract}

\newpage

\baselineskip=17pt

\section{Introduction}

Kaon physics has been a very fruitful field in particle physics since a long time.
This has been true in particular for the investigation of CP violation,
but also for the precise determination of fundamental parameters 
such as the Cabibbo angle $V_{us}$ or for the understanding of
low energy meson dynamics, for which perturbative QCD cannot be used and instead effective
theories as Chiral Perturbation Theory have to be applied.

Even though many results stem from experiments made in the 1970's,
thirty years later
the field of kaon decays is still of considerable interest for physics.
The reason for this is the huge amount of statistics accumulated by the main players
NA48 at CERN and KTeV at Fermilab, which makes it possible to do
precision measurements --- as e.g.\ for the determination of direct CP violation ---
as well as to search for extremely rare decays.

Both the NA48 and KTeV experiments were built for the precise determination of $\Reepe$,
which is a measure of the amount of direct CP violation in kaon decays.
The two experiments derive the beams of neutral kaons from
interactions of protons with 400~GeV (NA48) or 800~GeV (KTeV) energy on a fixed target.
Charged particles in the kaon beam are deflected by sweeping magnets.
To ensure pure $\kl$ beams, the neutral particles have to travel more than 100~m
before reaching the fiducial decay volume of similar length.
To create a beam of short-lived $\ks$ mesons the NA48 experiment uses
a second target, positioned slightly above the $\kl$ beam-line and shortly before
the decay volume. KTeV on the other hand uses
a regenerator placed in front of one of its two $\kl$ beam lines to
produce $\ks$ mesons.

The measurement of direct CP violation, described in the next section,
has been the original purpose of the two experiments.
However, either with the same data or with data from special data taking periods,
also the investigation of many rare $\kl$ and $\ks$ decays has been possible.
The most recent results are reported in the sections \ref{sec:kldecays} and \ref{sec:ksdecays}.
In the last section an outlook on measurements of $K^\pm$ decays with the NA48/2 experiment 
is given.

\section{Measurement of direct CP violation}
\label{sec:directcpv}

In 1964 Christenson, Cronin, Fitch, and Turlay
discovered CP violation in the decay of the long-lived kaon $\kl$ to two pions~\cite{bib:cpv64}.
It is explained by the small admixture of the eigenstates with opposite CP
to the mass eigenstates: 
$| \kl \ket \propto | K_2 \ket + \epsilon \: | K_1 \ket$ and 
$| \ks \ket \propto | K_1 \ket + \epsilon \: | K_2 \ket$,
with the CP eigenstates $| K_1 \ket$ and  $| K_2 \ket$ for $CP = +1$ and $-1$, resp.,
and the parameter $\epsilon$ measured to $|\epsilon| = (2.28 \pm 0.02) \times 10^{-3}$.
In the Standard Model, this so-called indirect CP violation is explained by 
the complex phase of the CKM matrix and mediated by $\KKbar$ oscillations
via box diagrams.

A second possibility of CP violation is the direct decay of the $K_2$ state to two pions.
In contrast to indirect CP violation it leads to different partial widths of the decays of $K^0$
and $\Kbar$ to two pions.
Its strength is given by the parameter $\epsilon'$, which is of the order $10^{-6}$ and
can be measured by the amplitude ratios
\begin{equation}
\eta_{+-} \: =  \: \frac{A(\klpipi)}{A(\kspipi)}     \: \simeq \: \epsilon + \epsilon'
\quad {\rm and} \quad
\eta_{00} \: =  \: \frac{A(\klpizpiz)}{A(\kspizpiz)} \: \simeq \: \epsilon - 2 \, \epsilon'.
\end{equation}
Within the Standard Model direct CP violation proceeds via penguin diagrams and is in general non-vanishing, 
while for other theories --- as e.g.\ a hypothetical super-weak
interaction --- $\epsilon'$ has to be zero.

Both experiments NA48 and KTeV measure direct CP violation via the double ratio
\begin{equation}
R \; \equiv  \; \frac{\Gamma (\klpizpiz)}{\Gamma (\kspizpiz)} \; \Big{/} \; \frac{\Gamma (\klpipi)}{\Gamma (\kspipi)}
        \; \approx \; 1 - 6 \times {\rm Re} \left( \frac{\epsilon'}{\epsilon} \right).
\end{equation}
This method has the advantage of the cancellation of many systematic uncertainties,
when all four decay modes are measured simultaneously.

The NA48 collaboration has recently finished the analysis on their complete data set,
which contains data from the years 1997--99 and 2001.
The data sample contains about $5 \times 10^6$ recorded events of the statistically limiting decay $\klpizpiz$.
The KTeV collaboration has so far published results from their 1996 and 97 data sets, which correspond to a total
of $3.4 \times 10^6$ reconstructed $\klpizpiz$ decays. The 1999 KTeV data are still being analyzed.

\begin{figure}[t]
\begin{minipage}{0.5\linewidth}
\vspace*{2mm}
\begin{center}
\begin{tabular}{l|r|r|}
                              & NA48      & KTeV      \\
                              & (2001)    & (96/97)   \\ \hline \hline
Trigger efficiency            & $\pm 0.6$ & $\pm 0.6$ \\
$\pi^+ \pi^-$ reconstruction  & $\pm 0.5$ & $\pm 0.3$ \\
$\pi^0 \pi^0$ reconstruction  & $\pm 0.9$ & $\pm 1.5$ \\
Background                    & $\pm 0.7$ & $\pm 1.1$ \\
Accidental activity           & $\pm 0.6$ & ---       \\
$\ks$ tagging                 & $\pm 0.7$ & ---       \\
Acceptance                    & $\pm 0.7$ & $\pm 0.9$ \\
MC statistics                 & $\pm 0.6$ & $\pm 0.6$ \\
Fit procedure                 & ---       & $\pm 0.3$ \\ \hline \hline
Total                         & $\pm 1.8$ & $\pm 2.4$ \\ \hline
\end{tabular}
\vspace*{3mm}
\captionof{table}{\it Systematic uncertainties on $\Reepe$ for NA48 and KTeV.}
\label{tab:reepe_syst}
\end{center}
\end{minipage}
\hfill
\begin{minipage}{0.45\linewidth}
\centerline{
\includegraphics[width=0.84\linewidth]{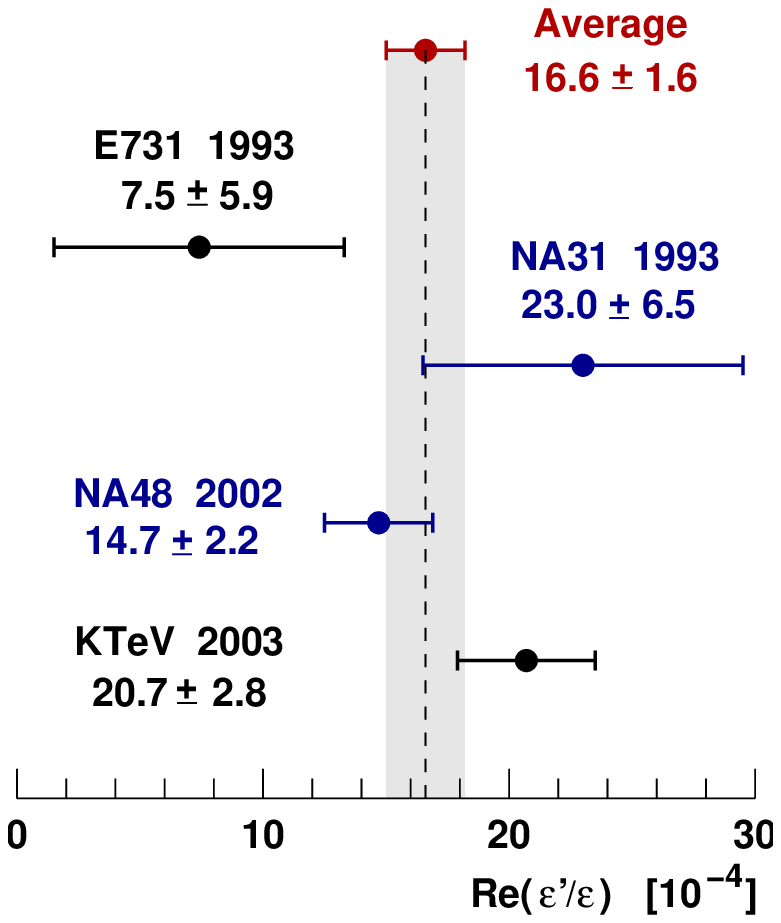}
}
\caption{\it Recent measurements and current world average for $\Reepe$.
\label{fig:reepe_results}}
\end{minipage}
\end{figure}

The analysis methods of the two experiments are fairly similar, 
with the exception of the treatment of $\ks$ and $\kl$ acceptances.
NA48 weights the $\kl \to 2 \pi$ events with the $\ks$ life-time in order to
reduce the $\ks$-$\kl$ acceptance differences to pure beam geometry. This method however
loses about $70\%$ of the $\kl$ statistics.
The KTeV collaboration on the other hand uses a Monte Carlo simulation 
to account for the acceptance differences, which allows the use of the whole
$\kl$ statistics for the analysis, but results in an about 20 times larger correction
on the result.
Both experiments have made huge efforts to understand and possibly 
reduce systematic uncertainties. A list of the systematics
is given in Tab.~\ref{tab:reepe_syst}.
The largest uncertainty comes from the energy scale and possible non-linearities
of the electromagnetic calorimeters used to measure the $K^0 \to 2 \pi^0$ decays.

The final result of the NA48 experiment on the combined 1997--99 and 2001 data is
$\Reepe \; = \; (14.7 \pm 1.4_{\subrm{stat}} \pm 1.7_{\subrm{syst}}) \times 10^{-4}$~\cite{bib:na48_reepe01}.
The KTeV collaboration has published the measurement
$\Reepe \; = \; (20.7 \pm 1.5_{\subrm{stat}} \pm 2.4_{\subrm{syst}}) \times 10^{-4}$
on the 1996 and 97 data~\cite{bib:KTeV_reepe97}.
The combined world average, including the results of the predecessor experiments
NA31 at CERN and E731 at Fermilab as shown in Fig.~\ref{fig:reepe_results}, now is
\begin{equation}
\Reepe \; = \; (16.6 \pm 1.6) \times 10^{-4}.
\end{equation}
This establishes a non-zero value of $\epsilon'$ and therewith the existence of direct CP violation
in neutral kaon decays at a $10 \, \sigma$ level.

However, the theoretical interpretation of the value of the result remains difficult, as long range
contributions dominate and the QCD and electro-weak penguin amplitudes destructively interfere.
Furthermore, a strong dependence on the strange quark mass exists.
A large number of groups have tried to predict $\Reepe$, but results vary by more of an order of magnitude
and still have large uncertainties.

\section{Rare $\kl$ decays}
\label{sec:kldecays}

Apart from the measurement of direct CP violation,
both experiments have accumulated large amounts of statistics of $\kl$ decays.
The KTeV collaboration has performed special data taking periods in 1997 and 1999
(experiment E799) for the investigation of rare $\kl$ decays.
This was not done in NA48, where rare $\kl$ decays were recorded together with
the $K \to 2 \pi$ decays used for the $\Reepe$ measurement.
The KTeV experiment has for most $\kl$ decays 
about $3-4$ times more statistics collected than NA48, they therefore dominate
most of the rare $\kl$ decay measurements.

\subsection{The decays $\kl \to \gamma^\star \gamma^\star$}

\begin{figure}[t]
\begin{minipage}{0.35\linewidth}
\vspace*{7mm}
\centerline{
\includegraphics[width=0.85\linewidth]{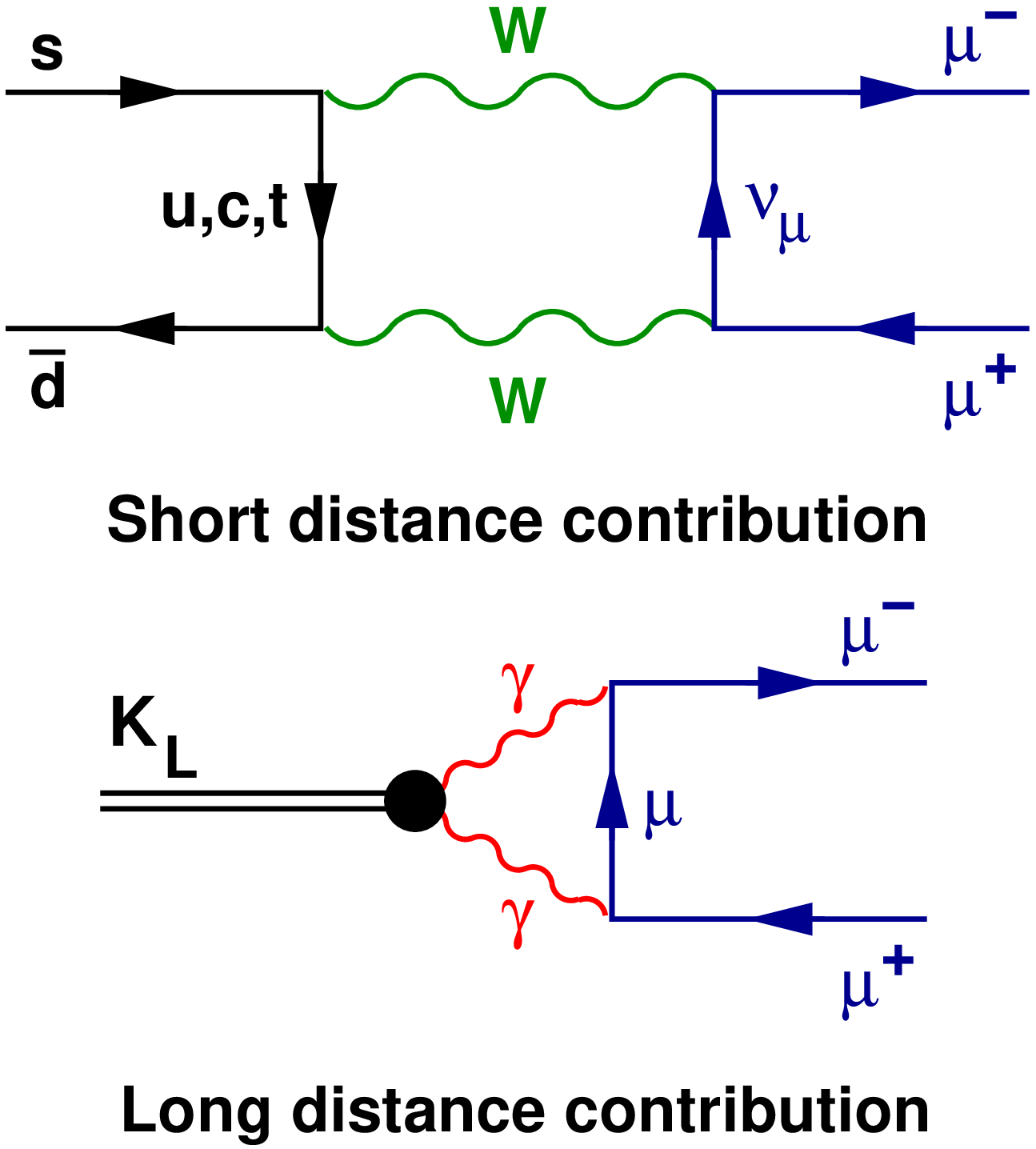}
}
\caption{\it Diagrams of short and long distance contributions to $\klmumu$.
\label{fig:klmumu_feynman}}
\end{minipage}
\hspace*{5mm}
\begin{minipage}{0.58\linewidth}
\centerline{
\includegraphics[width=0.9\linewidth]{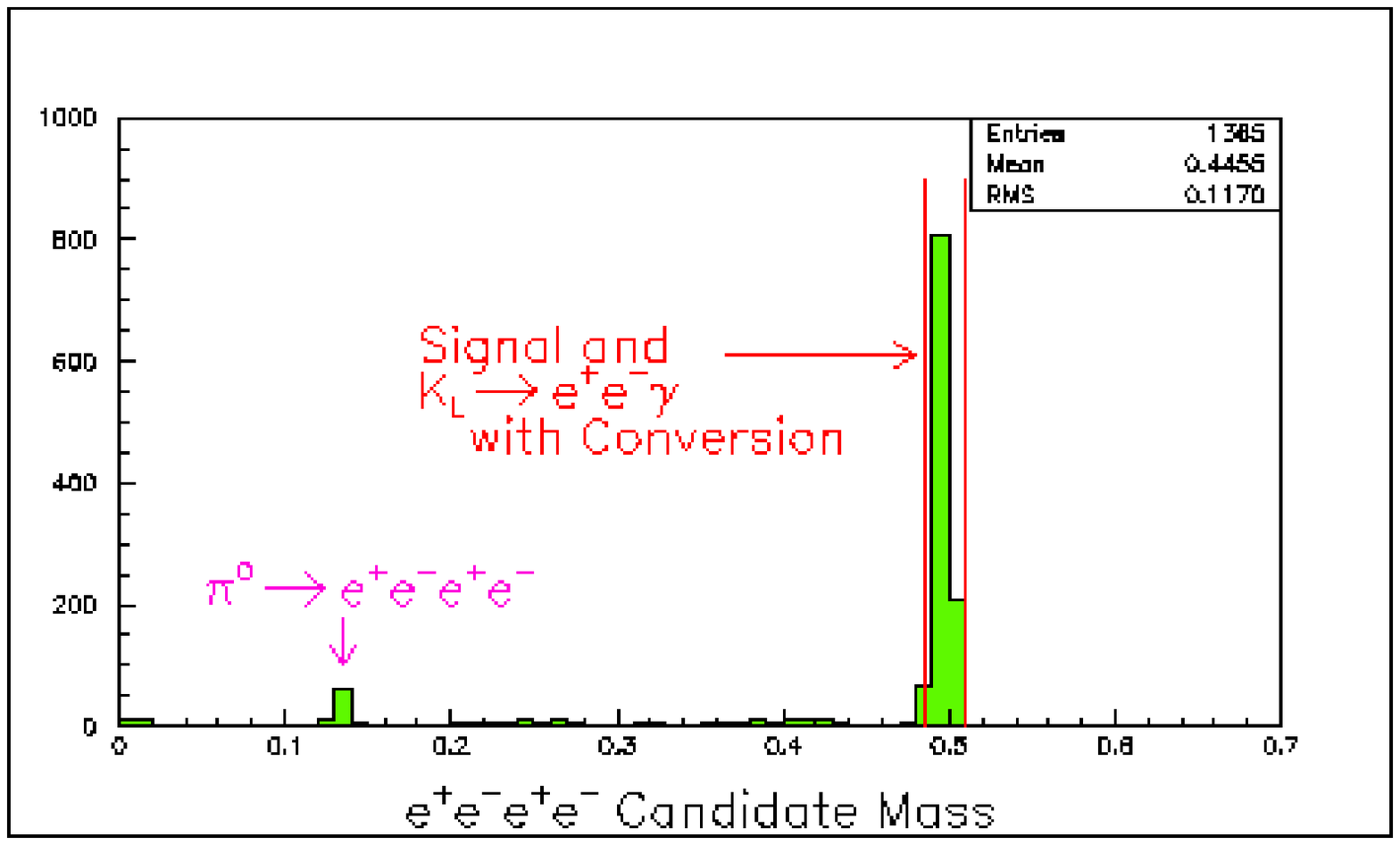}
}
\vspace*{4mm}
\caption{\it Invariant mass distribution of the KTeV $\kleeee$ candidates.
\label{fig:kleeee_mass}}
\end{minipage}
\end{figure}

Interest in the decays $\kl \to \gamma^\star \gamma^\star$
--- with the virtual photons either becoming real photons or producing
$e^+ e^- / \mu^+ \mu^-$ pairs --- is high, as they determine
the long distance contribution to the decay $\klmumu$ (Fig.~\ref{fig:klmumu_feynman}).
For this measurement the decays $\kleeee$ and $\klmumuee$ are well suited, however, 
they also exhibit very small branching fractions.

The KTeV collaboration has recently analyzed their full 1997 and 99 data set for
$\kleeee$ decays. They observe 1056 signal events, with a background expectation
of only 5~events, which come from $\kleeg$ decays with an external $\gamma$ conversion
on detector material (Fig.~\ref{fig:kleeee_mass}).
The preliminary result on the branching fraction is
\begin{equation}
\Br(\kleeee) = ( 4.07 \pm 0.12_{\subrm{stat}} \pm 0.11_{\subrm{syst}} \pm 0.16_{\subrm{norm}}) \times 10^{-8}.
\end{equation}
The evaluation of the form factor is in progress.

\begin{figure}[t]
\centerline{
\includegraphics[width=0.6\linewidth]{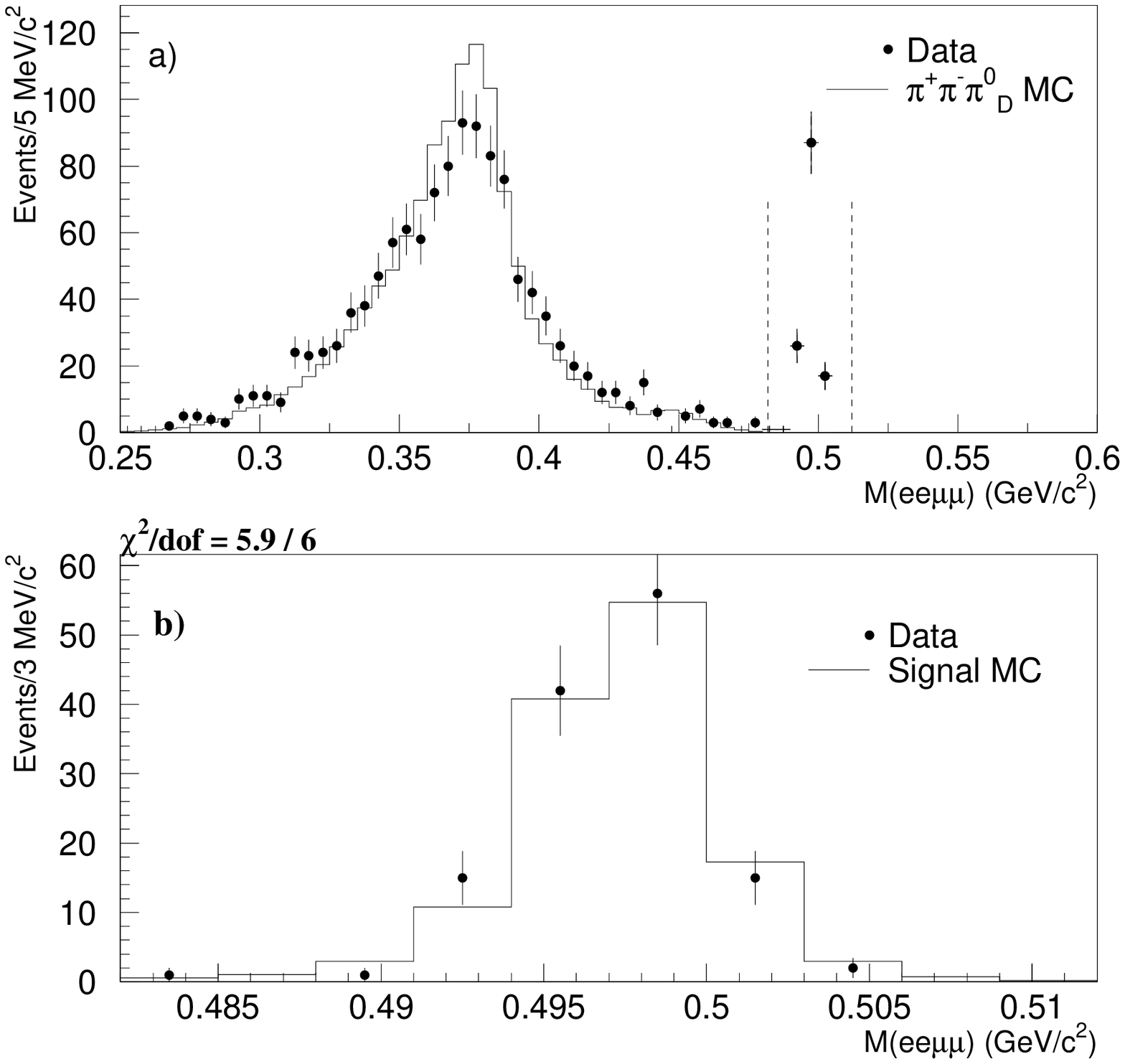}
}
\caption{\it Invariant mass distribution of the KTeV $\klmumuee$ candidates.
\label{fig:klmumuee_mass}}
\end{figure}

KTeV has also published a measurement of $\klmumuee$ decays based on their full data set.
They find 132 events in the signal region with a background expectation of 0.8~events (Fig.~\ref{fig:klmumuee_mass}),
which results in a branching fraction measurement~\cite{bib:ktev_klmumuee} of
\begin{equation}
\Br(\klmumuee) = ( 2.69 \pm 0.24_{\subrm{stat}} \pm 0.12_{\subrm{syst}}) \times 10^{-9}.
\end{equation}
The measurement of the form factor agrees with measurements of the other 
$\kl \to \gamma^\star \gamma^{(\star)}$ modes, but suffers from the
low statistics in this channel.

\subsection{Search for the decay $\klpiee$}
\label{sec:klpi0ee}

\begin{figure}[t]
\begin{minipage}{0.4\linewidth}
\vspace*{7mm}
\centerline{
\includegraphics[width=0.9\linewidth]{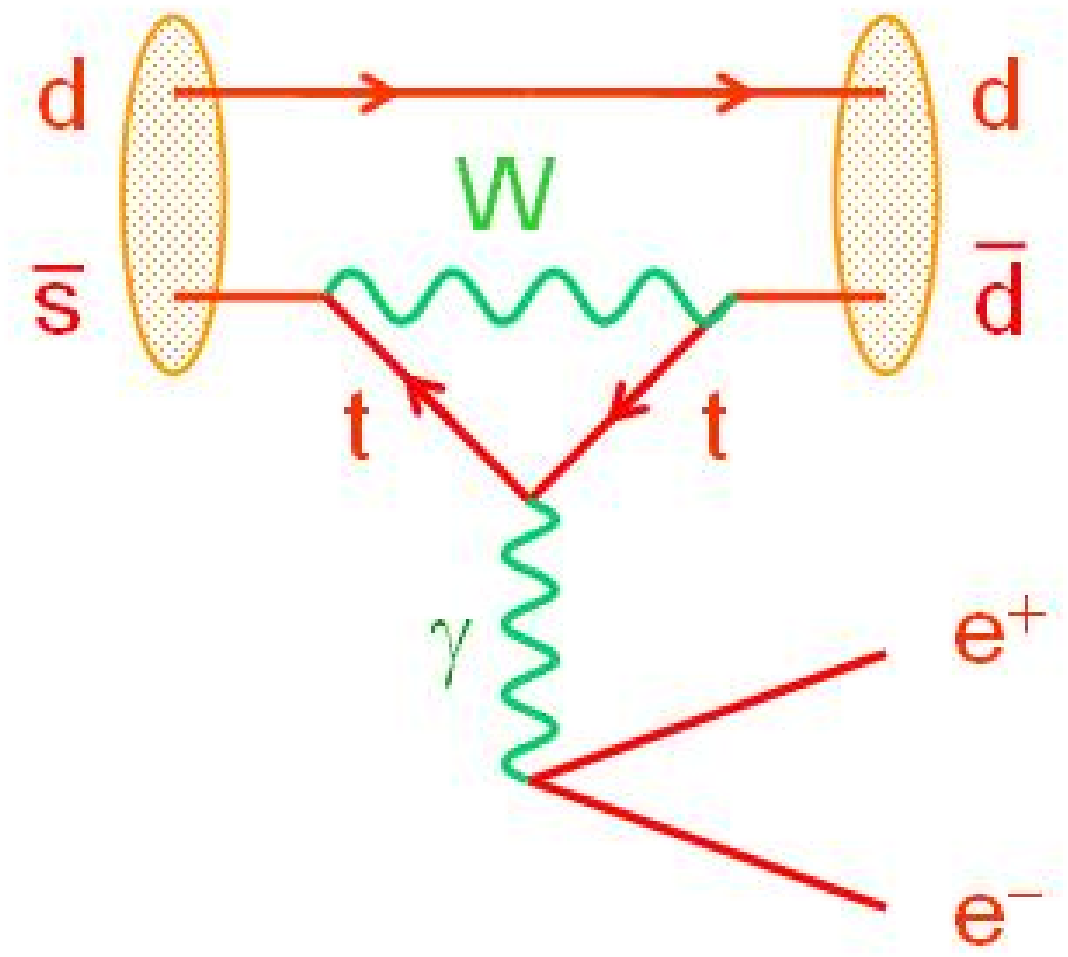}
}
\vspace*{3mm}
\centerline{
\includegraphics[width=0.9\linewidth]{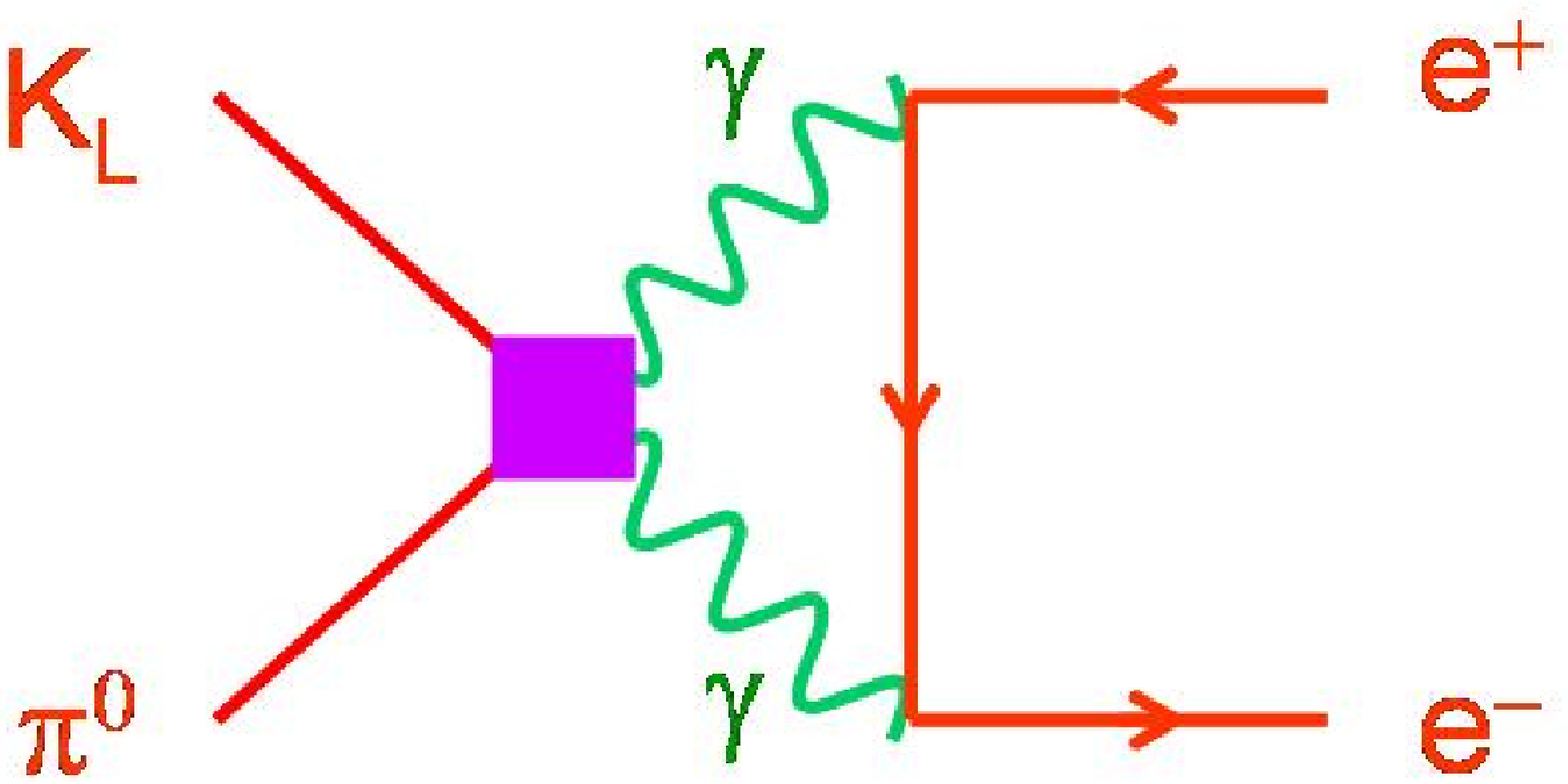}
}
\vspace*{2mm}
\caption{\it Direct CP violating (top) and CP conserving amplitude (bottom) of $\klpiee$.
\label{fig:klpiee_feynman}}
\end{minipage}
\hfill
\begin{minipage}{0.5\linewidth}
\centerline{
\includegraphics[width=0.95\linewidth]{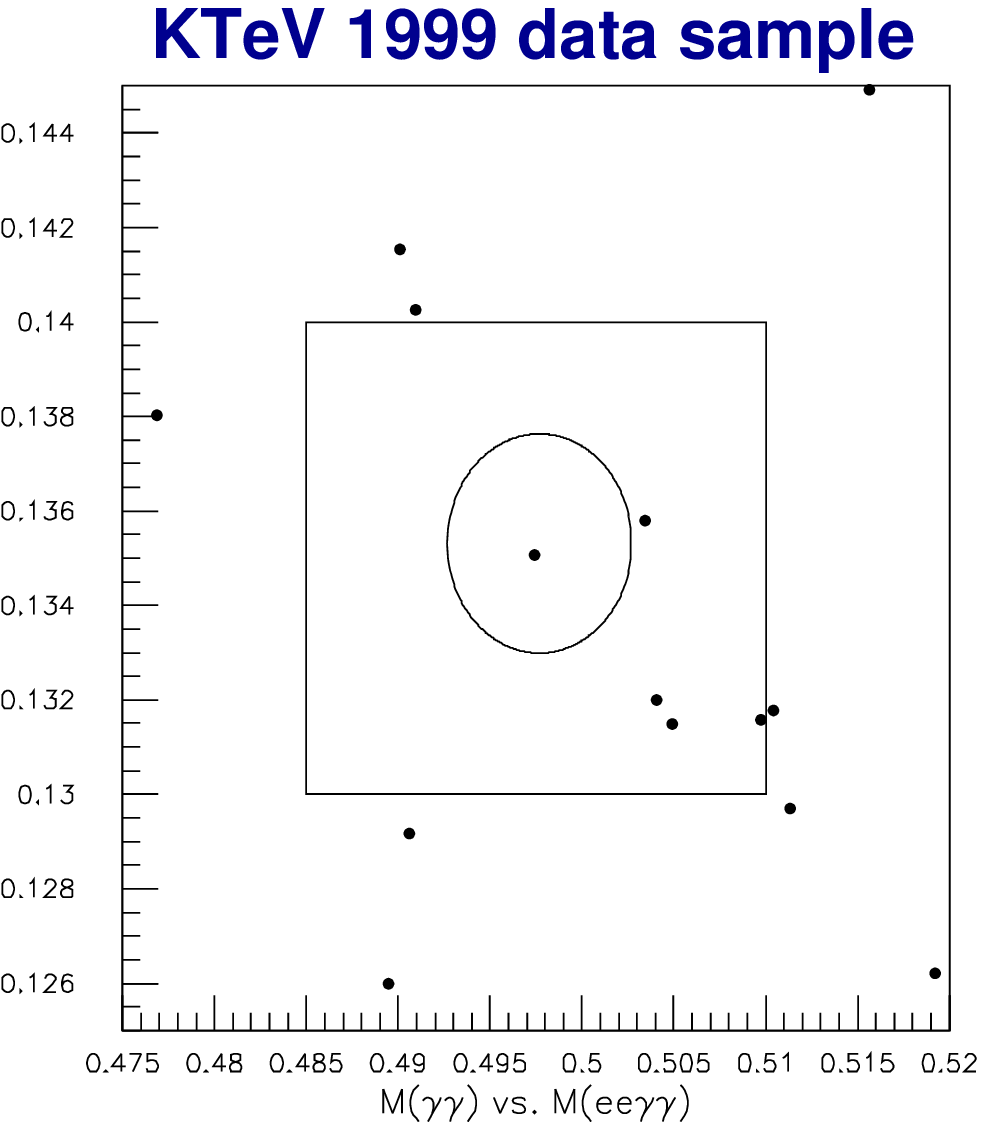}
}
\caption{\it Invariant $\gamma \gamma$ vs.\ $e^+ e^- \gamma \gamma$ mass for $\klpiee$ signal candidates in the KTeV 1999 data.
\label{fig:klpiee_signal}}
\end{minipage}
\end{figure}

The very rare decay $\klpiee$ is expected to have a sizable direct
CP violating amplitude from electro-weak penguin transitions (Fig.~\ref{fig:klpiee_feynman} top).
This amplitude is directly proportional to Im$(\lambda_t) = $Im$(V_{td} V_{ts}^\star) = \eta A^2 \lambda^5$. 
and allows therefore the measurement of height $\eta$ of the unitarity triangle. The corresponding contribution to the 
$\klpiee$ branching fraction has been estimated to a few $10^{-12}$.
However, also indirect CP violating and CP conserving amplitudes exist for $\klpiee$. The latter
can be deduced from the measurement of $\klpigg$ decays (see Fig.~\ref{fig:klpiee_feynman} bottom).
Its contribution to $\Br(\klpiee)$ has been determined to $(0.5 \pm 0.2) \times 10^{-12}$~\cite{bib:na48_klpigg}
and is small compared to the expected contribution from direct CP violation.
The contribution from indirect CP violation can be fixed by the measurement of $\kspiee$, which is discussed
in Sec.~\ref{sec:kspi0ee}.

The KTeV collaboration has used both their 97 and 99 data sets to search for $\klpiee$ events.
The main problem of the analysis is the rejection of the Greenlee background $\kleegg$,
which has a branching fraction of $6 \times 10^{-7}$ and may fake a $\klpiee$ decay if the photons accidentally
have a $\pi^0$ invariant mass.
This background is suppressed by using the topology of $\kleegg$ events, which arise
from $\kl \to \gamma^\star \gamma \to e^+ e^- \gamma$ decays with internal bremsstrahlung and therefore
have one photon emitted close to an electron track.

Applying all selection criteria, KTeV observes one event in the signal box
in the 99 data set (Fig.~\ref{fig:klpiee_signal}),
consistent with the background expectation of 0.99 events from $\kleegg$ decays,
which is converted into an upper limit of $\Br(\klpiee) < 3.5 \times 10^{-10}$
at $90\%$ confidence level~\cite{bib:ktev_klpiee99}.
Using also the published result on the 97 data set~\cite{bib:ktev_klpiee97}, this limit is improved to 
\begin{equation}
\Br(\klpiee) < 2.8 \times 10^{-10} \quad {\rm at} \: 90\% \: {\rm CL,}
\end{equation}
which is still about two orders of magnitude above the theoretical expectation
for the direct CP violating amplitude.


\section{Rare $\ks$ decays}
\label{sec:ksdecays}

In $\epsilon'$ data taking, only moderate $\ks$ intensity is needed, since 
the $\kl \to 2 \pi$ decays limit the overall statistics, while 
the decay of $\ks$ into $2 \pi$ is dominant. 
The investigation of rare $\ks$ decays therefore requires a different set-up as for
the $\Reepe$ measurement.

In 2002, the NA48 successor experiment NA48/1~\cite{bib:na48_1} has performed 
a special high-intensity run period with no $\kl$ beam but with about 200 times 
increased proton intensity on the $\ks$ target w.r.t.\ normal $\epsilon'$ data taking.
While the detector itself had only to be slightly modified,
the read-out had to be improved and partly re-done to deal with
the much higher data rate.

In addition, the NA48 experiment has performed a high-intensity $\ks$ run period in the year 2000.
As the drift chambers were not operational in 2000, this run has been used for investigating
$\ks$ decays into neutral final states as $\kspipipi$, $\ksgg$, and $\kspigg$.

\subsection{First observation of $\kspiee$}
\label{sec:kspi0ee}

\begin{figure}[t]
\centerline{\hbox{
\includegraphics[width=0.45\linewidth]{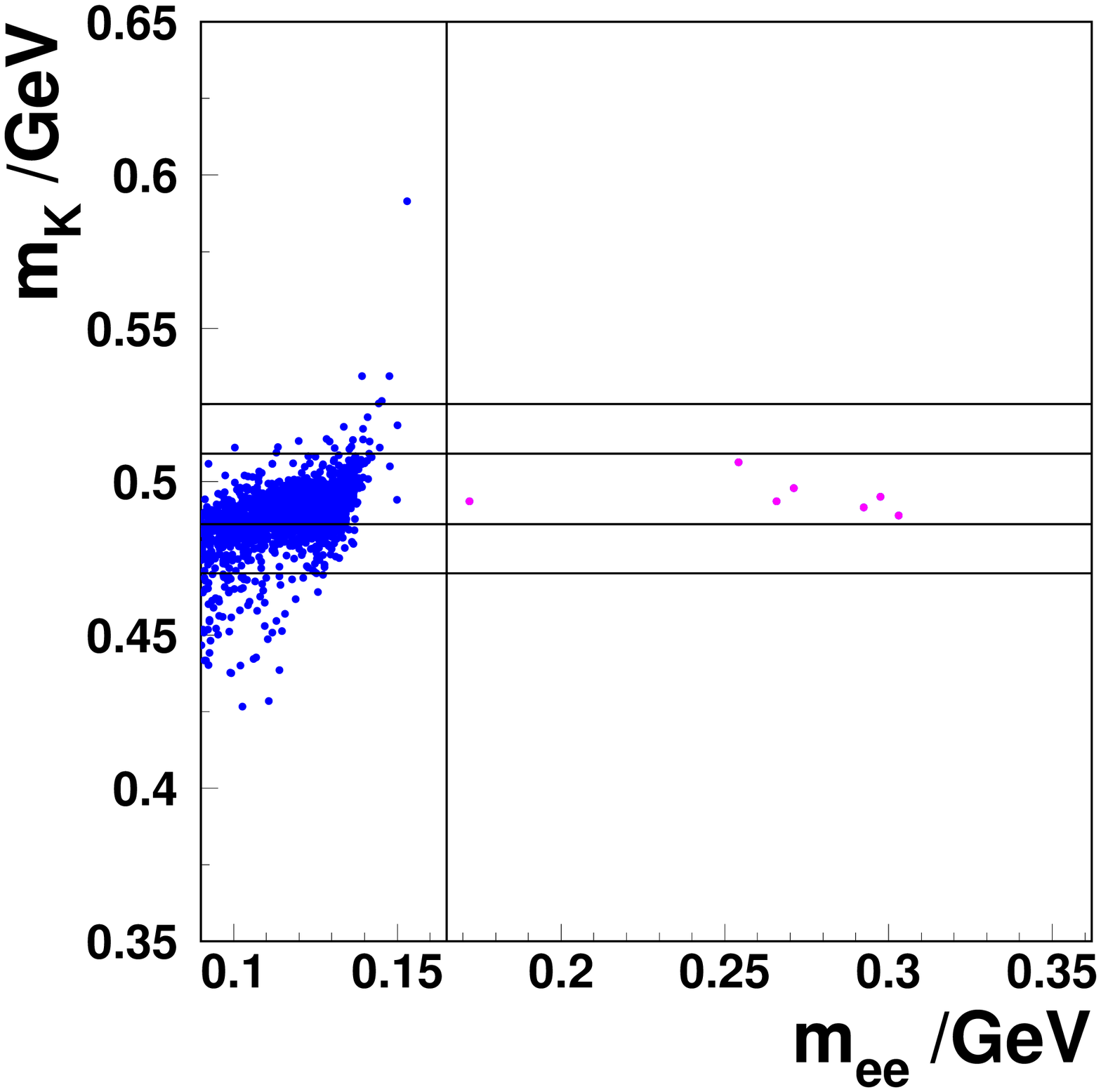}
\hspace*{8mm}
\includegraphics[width=0.45\linewidth]{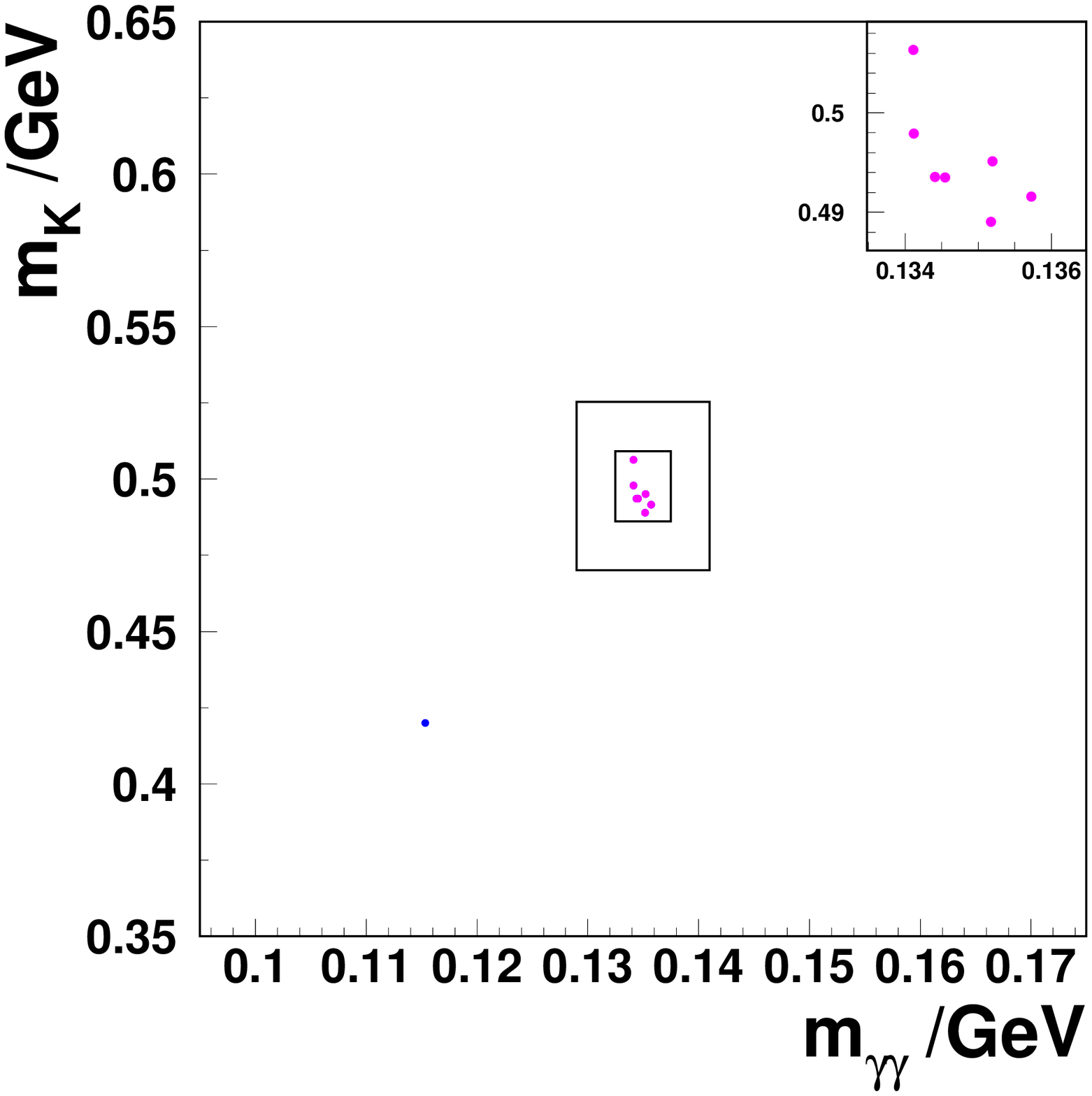}
}}
\caption{\it Distributions of invariant $\kspiee$ mass versus the $e^+e^-$ invariant mass (left) and
the two-photon invariant mass (right) in the NA48 data. 
Indicated are the borders of the signal and control regions and the cut
at $m_{ee} > 165$~MeV/$c^2$.
\label{fig:kspiee_signal}}
\end{figure}

\begin{figure}[t]
\centerline{
\includegraphics[width=0.44\linewidth]{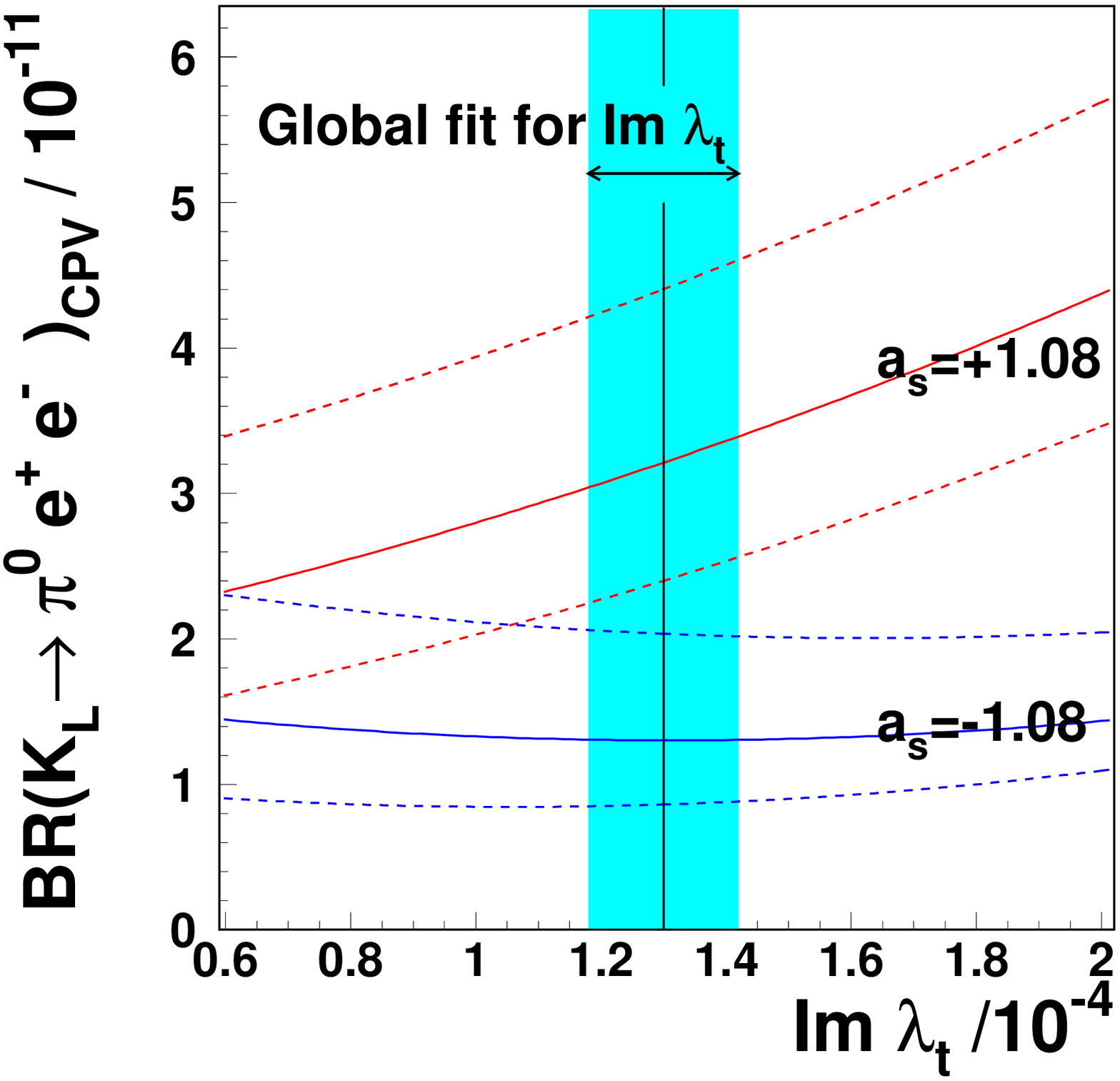}
}
\caption{\it Prediction for the CP violating fraction of the decay $\klpiee$ as function
of Im$(\lambda_t)$.
\label{fig:klpiee_prediction}}
\end{figure}

As described in the Section~\ref{sec:klpi0ee}, the decay
$\kspiee$ directly measures the indirect CP violating part 
of the corresponding $\kl$ transition.
The decay had not yet been observed, the best upper limit was
$\Br(\kspiee) < 1.4 \times 10^{-7}$,
determined by the NA48 experiment using a short test $\ks$ data taking period in 1999~\cite{bib:na48_kspiee99}.
A similar analysis has now been performed on the much higher statistics of the 2002 NA48 data set.
To suppress $\ks \to \pi^0 \pid$ decays (with $\pid \to e^+ e^- \gamma$),
an invariant $e^+ e^-$ mass above 165~MeV/$c^2$ was required for the signal candidates.
The analysis was performed blind with both the signal and control region masked.
After opening the signal box, 7 signal candidates were found with an estimated background of
0.15~events from mainly $\kleegg$ and overlapping events (Fig.~\ref{fig:kspiee_signal}).
This is the first observation of this decay, and corresponds to a (preliminary) branching fraction of 
\begin{equation}
\Br(\kspiee)_{m_{ee}>165 \: {\rm MeV}} = (3.0 \, {+1.5 \atop -1.2}_{\subrm{stat}} \, \pm 0.2_{\subrm{syst}}) \times 10^{-9}.
\end{equation}
Using the matrix element calculated by D'Ambrosio {\em et al.}~\cite{bib:dambrosio} with the form factor set to 1,
this turns into an overall branching fraction of
$\Br(\kspiee) = (5.8 {+2.8 \atop -2.3}_{\subrm{stat}} \pm 0.3_{\subrm{syst}} \pm 0.8_{\subrm{theo}}) \times 10^{-9}$.
Finally, using this result for the prediction of the CP violating $\klpiee$ amplitudes, it follows
\begin{equation}
\Br(\klpiee)_{\subrm{CPV}} = ( 17.7_{\subrm{indirect CPV}} \pm  9.5_{\subrm{interference}} + 4.7_{\subrm{direct CPV}} ) \times 10^{-12},
\end{equation}
which shows the dominance of the indirect CP violating amplitude of $\klpiee$ and the difficulty to extract
Im$\lambda_t$ from a future $\klpiee$ measurement (Fig.~\ref{fig:klpiee_prediction}).

\subsection{Search for CP violation in $\kspipipi$}

The decay $\kspipipi$ is purely CP violating in complete analogy, but with reversed CP values, 
to $\klpizpiz$.
The amplitude ratio compared to the corresponding $\kl$ decay is expected to be 
\begin{equation}
\etazzz \; \equiv \; \frac{A(\kspipipi)}{A(\klpipipi)}
        \; = \; \epsilon + i \, \frac{{\rm Im}(A_1)}{{\rm Re}(A_1)}.
\end{equation}
While the real part is fixed by CPT conservation
the imaginary part depends on the isospin 1 amplitude $A_1$ and may differ from 
Im$(\epsilon)$.
Experimentally, $\kspipipi$ has never been observed. The parameter $\etazzz$
has been measured to $\reeta = 0.18 \pm 0.15$ and $\imeta = 0.15 \pm 0.20$
by CPLEAR~\cite{bib:ks3pi0_cplear}. In addition, the SND experiment
has set a limit on the branching fraction of  $1.4 \times 10^{-5}$
at 90\% confidence level~\cite{bib:ks3pi0_snd}.

In NA48 $\kl$ and $\ks$ mesons are produced in equal amounts at the target. The dependence
of the $K^0/\overline{K^0} \to 3 \pi^0$ intensity as function of proper time therefore is given by
\begin{equation}
\begin{array}{rcl}
I_{3\pi^0}(t) & \propto & \underbrace{\raisebox{0mm}[0mm][2mm]{$e^{- \Gamma_L t}$}}_{\textstyle \raisebox{-1mm}{$\kl$ decay}} \! \! 
                    + \: \underbrace{\raisebox{0mm}[0mm][2mm]{$|\etazzz|^2 \, e^{- \Gamma_S  t}$}}_{\textstyle \raisebox{-1mm}{$\ks$ decay}} \\*[13mm]
        & &           \: + \: \underbrace{\raisebox{0mm}[0mm][2mm]{$2 \, D(p) \left( {\rm Re}(\etazzz) \cos{\Delta m  t}  -
                                                        {\rm Im}(\etazzz) \sin{\Delta m t} \right)
                                             \, e^{ - \frac{1}{2}(\Gamma_S + \Gamma_L) \, t}$}}_{\textstyle \raisebox{-1mm}{$\kl$-$\ks$ interference}}
\end{array}
\end{equation}
with the momentum dependent dilution
$D(p) = (N(K^0) - N(\overline{K^0})/(N(K^0) - N(\overline{K^0})$.
In NA48, this production asymmetry is on average about 0.35 with an almost linear
dependency from the kaon momentum.
While it is hopeless to observe the pure $\ks$ contribution above the $\kl$ decay,
the $\kl \ks$ interference is suppressed by only the first order in $\etazzz$.

\begin{figure}[t]
\centerline{\hbox{
\includegraphics[width=0.38\linewidth]{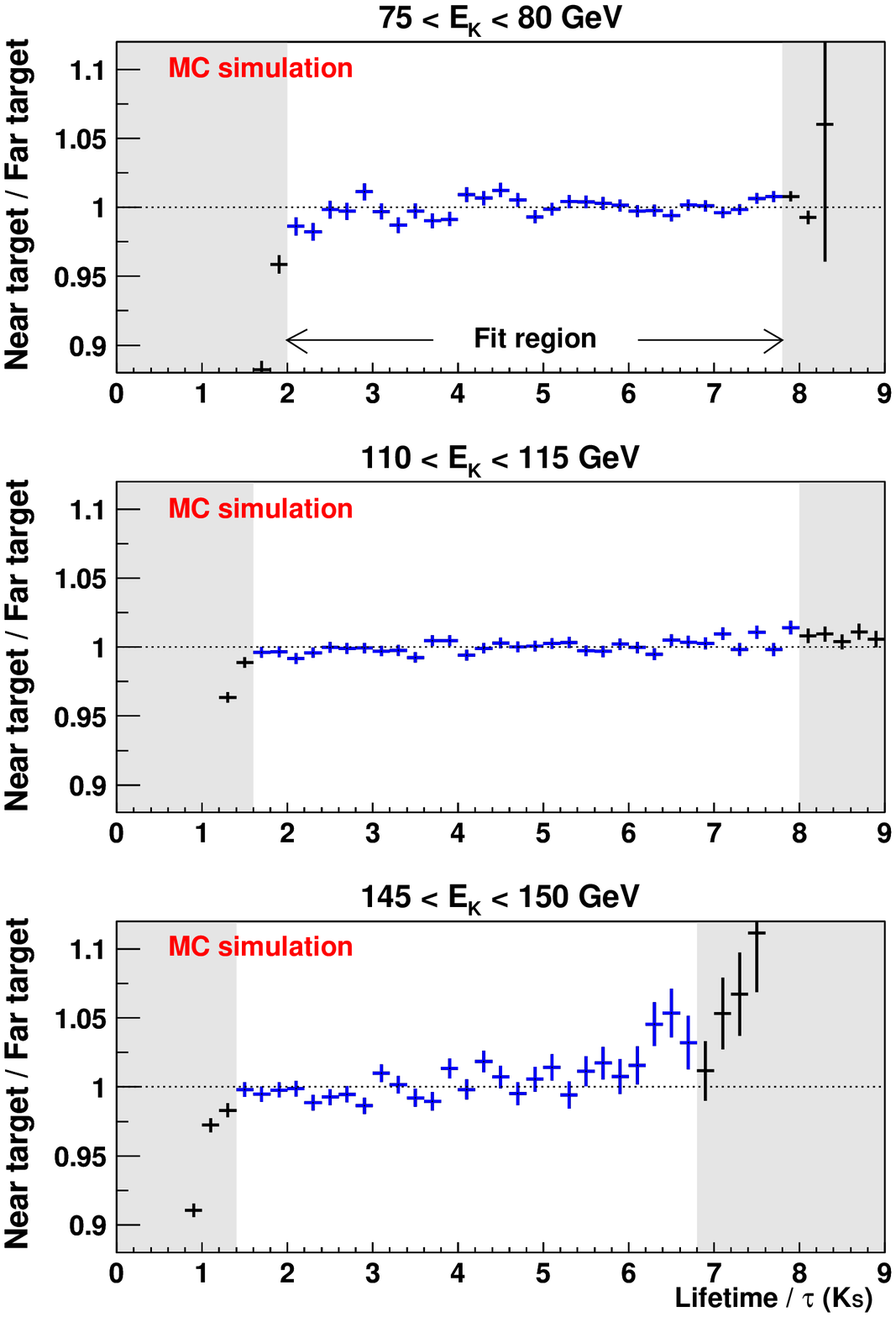}
\hspace*{10mm}
\includegraphics[width=0.38\linewidth]{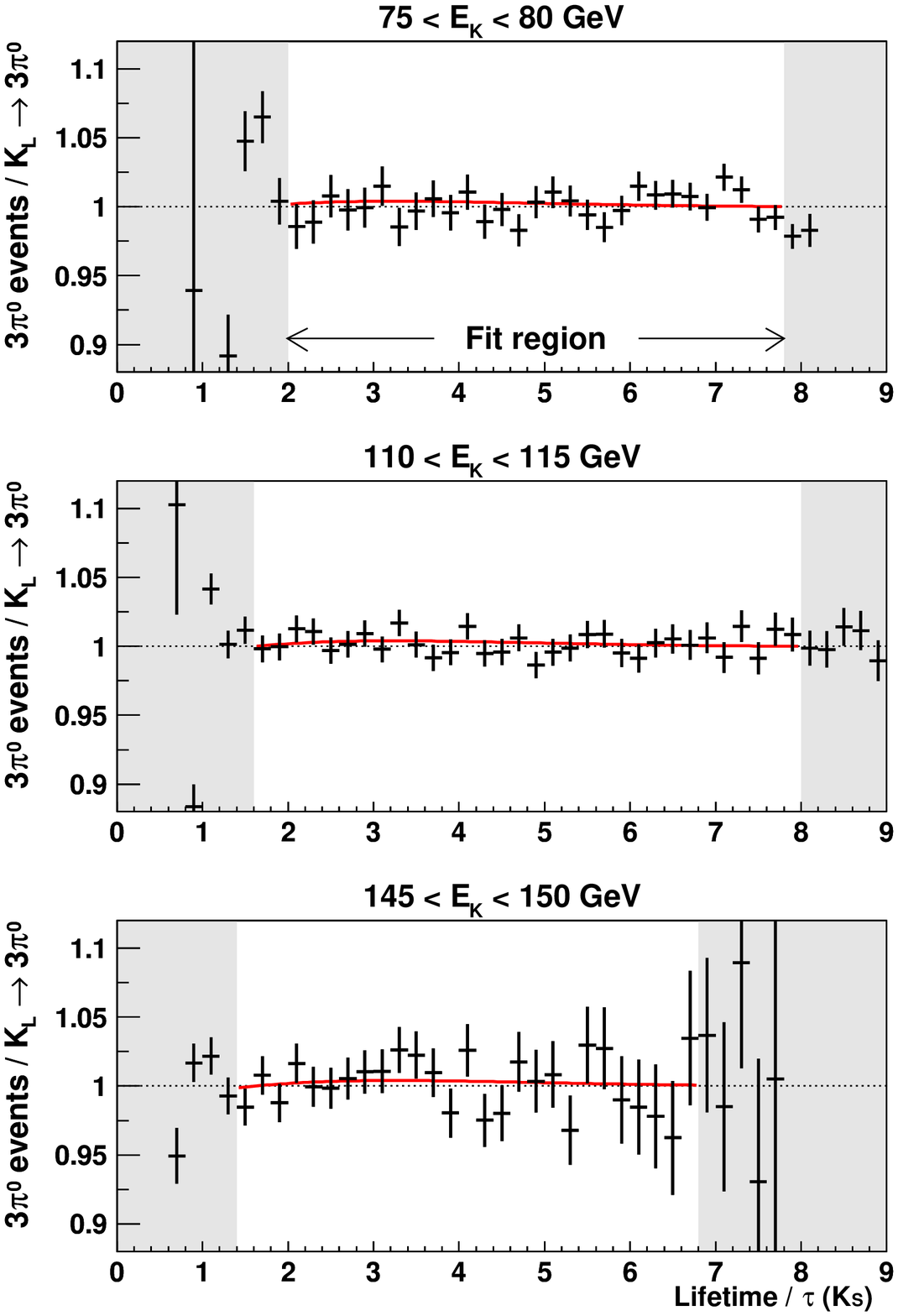}
}}
\caption{\it Ratios of $3 \pi^0$ events for different energy bins. Left: 
$\klpipipi$ Monte Carlo simulation from near and far target.
Right: Near-target data normalized to geometry-corrected far-target
$\klpipipi$ data.
\label{fig:ks3pi0_fit}}
\end{figure}

\begin{figure}[t]
\begin{minipage}{0.51\linewidth}
\begin{center}
\vspace{11mm}
\renewcommand{\arraystretch}{1.1}
\begin{tabular}{l|c|c}
                              & $\reeta$       & $\imeta$       \\ \hline
Acceptance                    & $\pm \, 0.003$ & $\pm \, 0.008$ \\
Accidental activity           & $\pm \, 0.001$ & $\pm \, 0.006$ \\
Energy scale                  & $\pm \, 0.001$ & $\pm \, 0.001$ \\
$K^0 \overline{K^0}$ dilution & $\pm \, 0.003$ & $\pm \, 0.004$ \\
Fit                           & $\pm \, 0.001$ & $\pm \, 0.002$ \\ \hline \hline
Total:                        & $\pm \, 0.005$ & $\pm \, 0.011$ \\
\end{tabular}
\renewcommand{\arraystretch}{1.0}
\vspace{5mm}
\captionof{table}{\it Systematic uncertainties on $\etazzz$.}
\label{tab:ks3pi0_syst}
\end{center}
\end{minipage}
\hfill
\begin{minipage}{0.45\linewidth}
\includegraphics[width=0.96\linewidth]{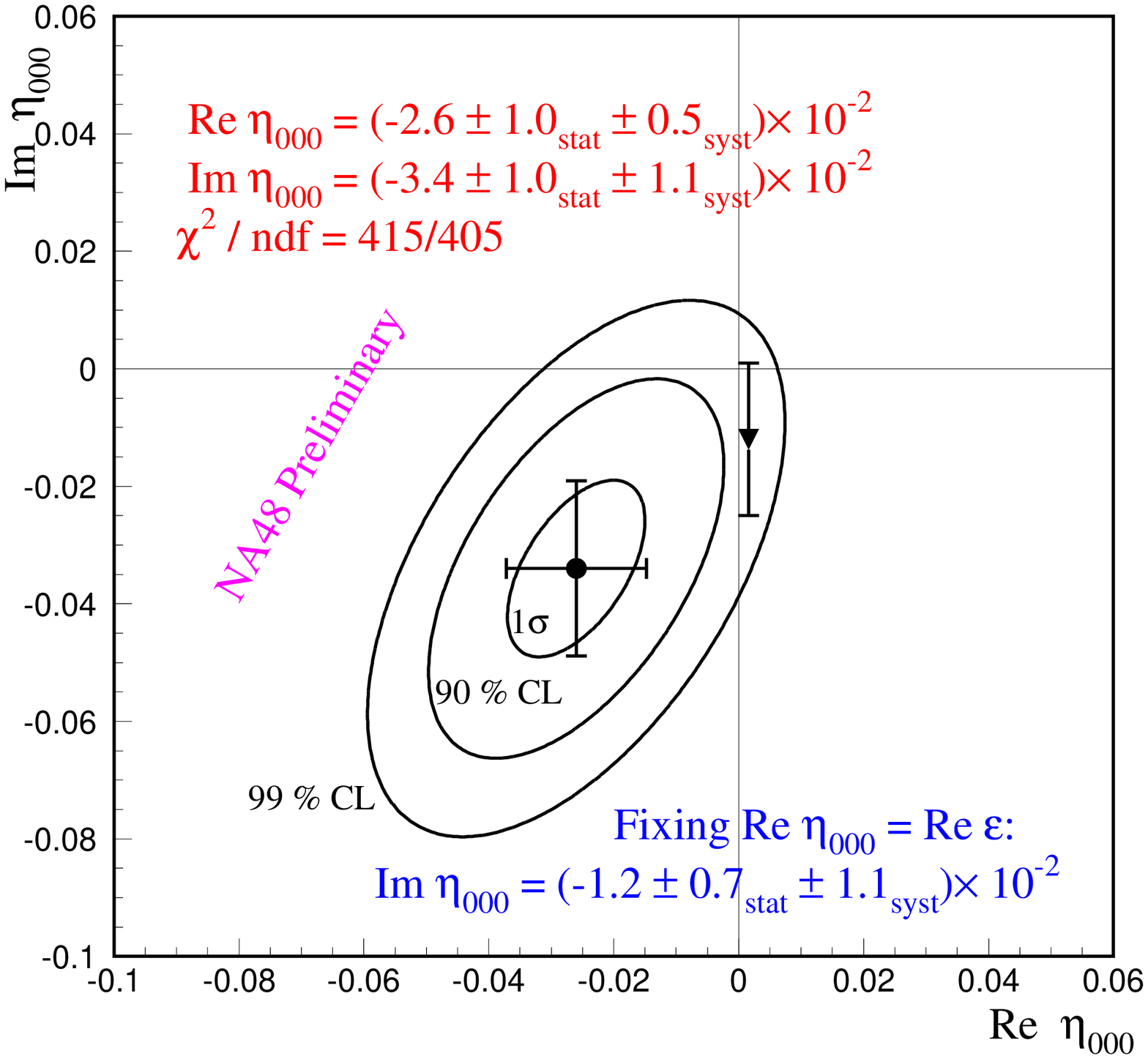}
\vspace{-4mm}
\caption{\it Fit result for $\etazzz$ (prel).
\vspace{-2mm}
\label{fig:ks3pi0_result}}
\end{minipage}
\end{figure}

For the $\etazzz$ measurement data from the 2000 near-target
run period were used. From these data about $6.5 \times 10^6$ $3\pi^0$ events were
selected with practically negligible background. 
To be as independent from Monte Carlo simulation as possible data from the far-target
run period of the same year were used for normalization to $\klpipipi$.
In this way, due to the almost identical geometry of near- and far-target beams, only residual effects had to be corrected for by Monte Carlo simulation
(see Fig.~\ref{fig:ks3pi0_fit} left).
To also be independent of the correct modeling of the different kaon energy spectra in the simulation,
the fit to the proper time distribution of the $3 \pi^0$ events was performed in bins of energy,
leaving all normalizations free in the fit.
The fit result is $\reeta = -0.026 \pm 0.010$ and $\imeta = -0.034 \pm 0.010$
with a correlation coefficient of 0.8.
The systematic uncertainties are dominated by uncertainties in the detector acceptance,
the accidental activity, and the $K^0 \overline{K^0}$ dilution (see Tab.~\ref{tab:ks3pi0_syst}).
The complete result, which is still preliminary, then is
\begin{equation}
\begin{array}{rcl}
\reeta & = & -0.026 \: \pm \: 0.010_{\subrm{stat}} \: \pm \: 0.005_{\subrm{syst}} \quad \mathrm{and} \quad \\
\imeta & = & -0.034 \: \pm \: 0.010_{\subrm{stat}} \: \pm \: 0.011_{\subrm{syst}}.
\end{array}
\end{equation}

The corresponding confidence limits are shown in Fig.~\ref{fig:ks3pi0_result}.
The result is consistent with 0 with roughly 5\% probability.
Turning this result into an upper limit on the branching fraction, one gets
\begin{equation}
\Br(\kspipipi) < 1.4 \times 10^{-6} \quad {\rm at \: 90\% \: CL,}
\end{equation}
which is one order of magnitude below the previous best limit.

Assuming CPT conservation, which fixes $\reeta$ to
${\rm{Re(}\epsilon\rm{)}} = 1.6 \times 10^{-3}$, one receives
\begin{equation}
\imeta|_{{\rm Re}(\etazzz) = {\rm Re}(\epsilon)} = -0.012 \pm 0.007_{\subrm{stat}} \pm 0.011_{\subrm{syst}}
\end{equation}
and $\Br(\kspipipi)|_{{\rm Re}(\etazzz) = {\rm Re}(\epsilon)} < 3.0 \times 10^{-7}$ at 90\% CL.

Finally, this result improves the limit on CPT violation via the
Bell-Steinberger relation. Using unitarity, this relation connects the CPT violating phase
$\delta$ with the CP violating amplitudes of the various $\kl$ and $\ks$ decays~\cite{bib:bellsteinberger}.
So far, the limiting quantity has been the precision of $\etazzz$. This new result, added to the measurements
of the other $\eta$ parameters~\cite{bib:pdg,bib:k3pi_cplear,bib:cpt_cplear},
improves the accuracy on Im$(\delta)$ by about 40\% to Im$(\delta) = (-1.2 \pm 3.0) \times 10^{-5}$,
now limited by the knowledge of $\eta_{+-}$.
Assuming CPT conservation in the decay, this can be converted to a new limit on the
$K^0\overline{K^0}$ mass difference 
of $m_{K^0} - m_{\overline{K^0}} = (-1.7 \pm 4.2) \times 10^{-19}$~GeV/$c^2$.

\subsection{Tests of Chiral Perturbation Theory in $\ksgg$ and $\kspigg$}

The neutral $\ks$ decays into the $\gamma \gamma$ and $\pi^0 \gamma \gamma$
final states are suited for investigating higher order predictions of Chiral Perturbation Theory (ChPT),
as contributions from lowest order ${\cal O}(p^2)$ do not exist.
For the decay $\ksgg$ an exact ${\cal O}(p^4)$ calculation predicts
the branching fraction to be $2.1 \times 10^{-6}$~\cite{bib:ksgg_theory}.

\begin{figure}[t]
\begin{minipage}{0.48\linewidth}
\centerline{\includegraphics[width=0.9\linewidth]{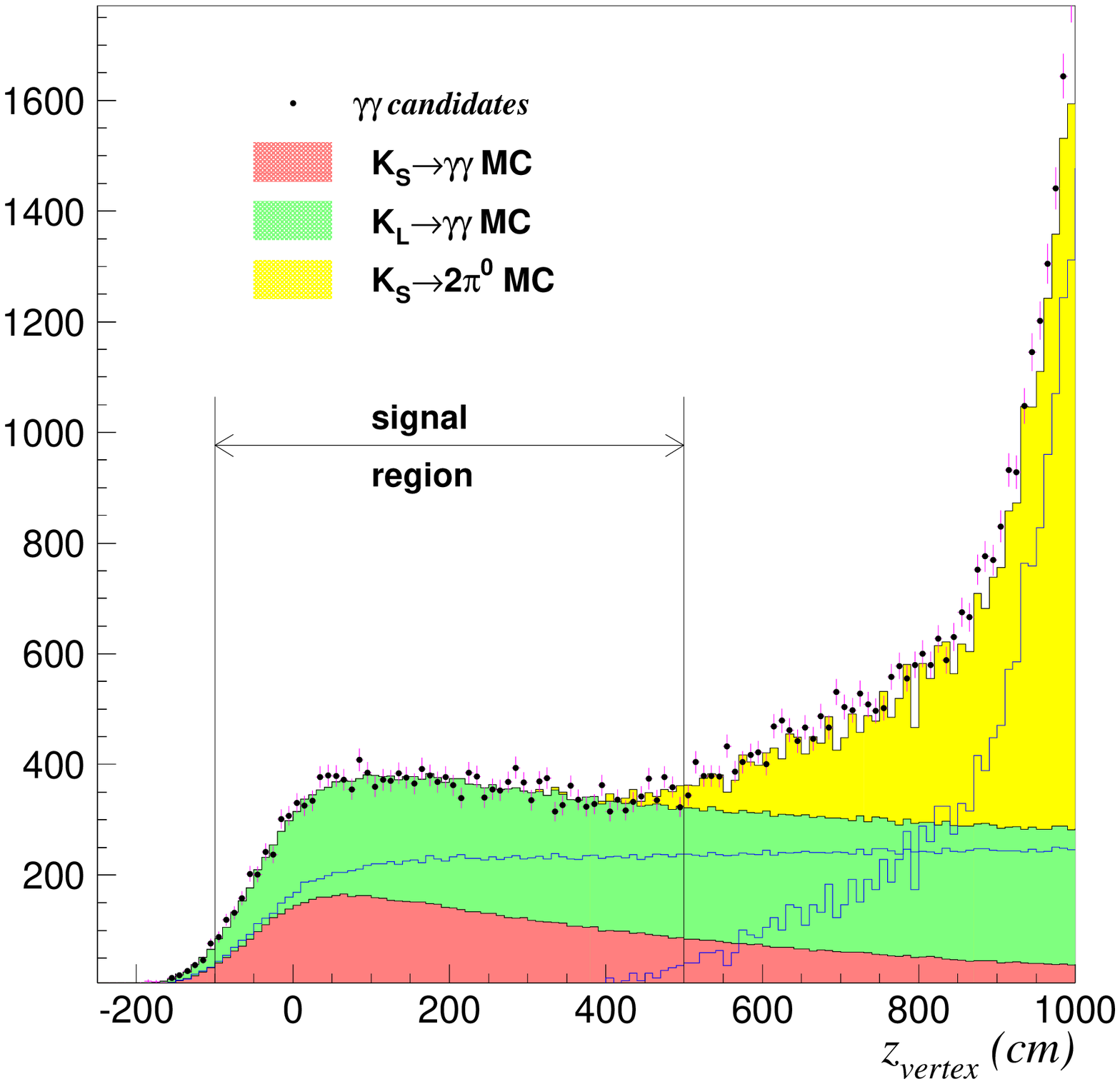}}
\caption{\it Longitudinal vertex distribution of the NA48 $\ksgg$ candidates.
\label{fig:ksgg}}
\end{minipage}
\hfill
\begin{minipage}{0.48\linewidth}
\centerline{\includegraphics[width=0.9\linewidth]{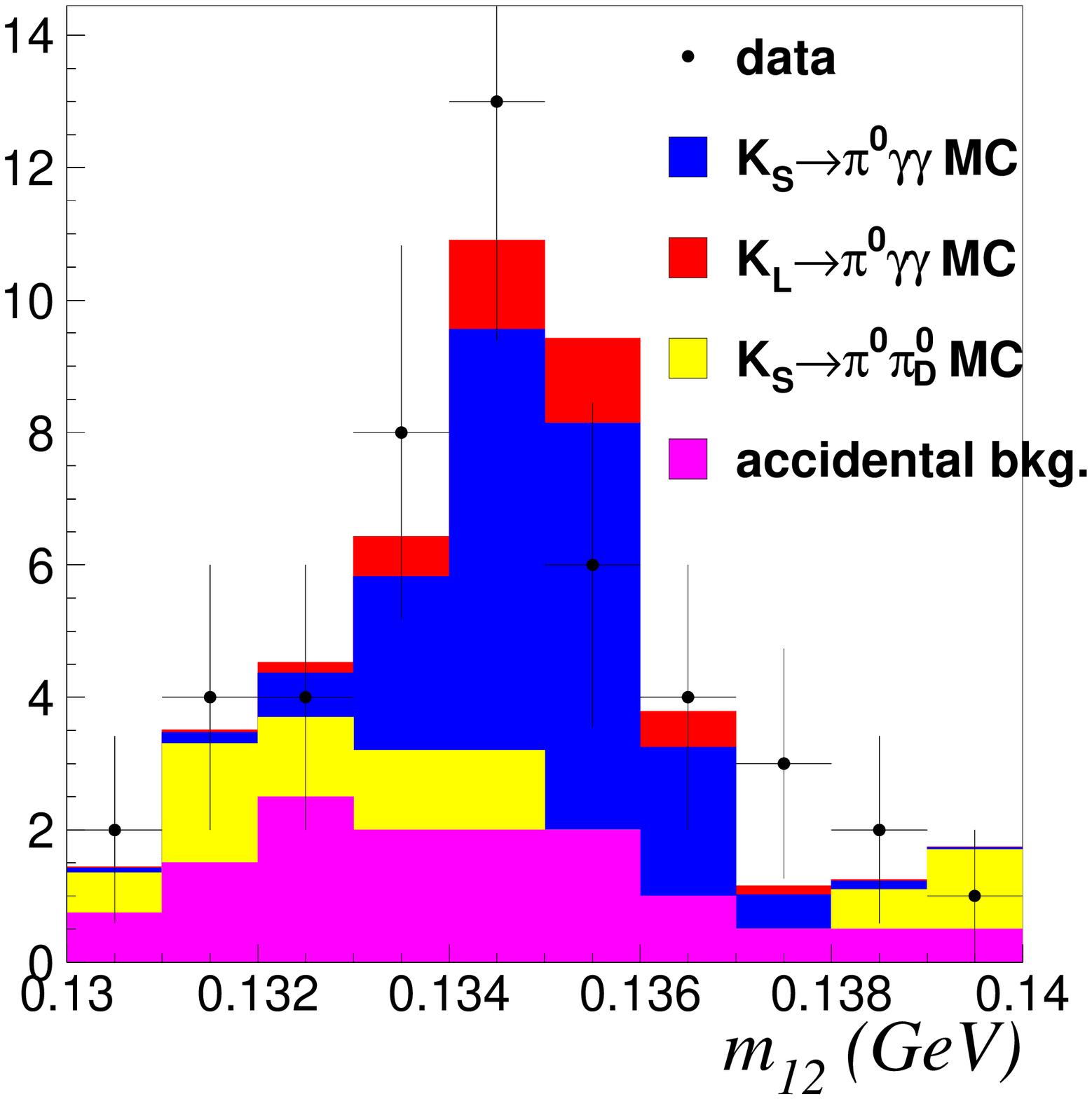}}
\vspace{-4mm}
\caption{\it Invariant $\pi^0 \to \gamma \gamma$ mass of the NA48 $\kspigg$ candidates.
\label{fig:kspigg}}
\end{minipage}
\end{figure}

NA48 has used the data from its 2000 high-intensity $\ks$ run period to investigate these decays.
For $\klsgg$ about 20000 candidate events are observed in the signal region between
$-1$ and 5~m from the final collimator (Fig.~\ref{fig:ksgg}).
About one third of these events originate from $\ks$ decays.
The corresponding branching fraction has been determined to~\cite{bib:ksgg}
\begin{equation}
\Br(\ksgg)= (2.78 \pm 0.06_{\subrm{stat}} \pm 0.04_{\subrm{syst}}) \times 10^{-6}.
\end{equation} 
This can be compared to the ${\cal O}(p^4)$ ChPT prediction given above,
showing a $\approx 30\%$ contribution from ${\cal O}(p^6)$ to the branching fraction.

For the decay $\kspigg$ the theoretical prediction using ChPT is
$\Br = 3.8 \times 10^{-8}$ for $z = m_{\gamma\gamma}^2/m_K^2 > 0.2$ to avoid the $\pi^0$ pole~\cite{bib:kspigg_theory}.
Analyzing its 2000 data set, the NA48 collaboration has found 31 signal candidates with an estimated
background of $13.6\pm2.8$ events from various sources (Fig.~\ref{fig:kspigg}).
From this, the branching fraction was calculated to~\cite{bib:kspigg}
\begin{equation}
\Br(\kspigg)|_{z>0.2}= (4.9 \pm 1.6_{\subrm{stat}} \pm 0.9_{\subrm{syst}}) \times 10^{-8}.
\end{equation}
This is the first observation of this decay. The measured branching fraction is in agreement with
the theory prediction, albeit the statistical precision does not yet 
allow concrete conclusions.

\section{Outlook on $K^\pm$ decays}
\label{sec:kcdecays}

In the year 2003 the NA48/2 experiment has performed a three month high intensity
data taking period for the investigation of $K^\pm$ decays.
For this special run the NA48 beam-line has been altered to simultaneously 
select $K^+$ and $K^-$ mesons with a momentum of about 60~GeV/$c$.
In addition, micro-mesh gas chambers have been installed
as beam spectrometer to achieve a $K^\pm$ momentum resolution of $\sim 1\%$.
Expected are about $3 \times 10^{11}$ $K^\pm$ decays in the fiducial volume
in the 2003 run period~\cite{bib:na48_2}.

The main interest of this run is the search for direct CP violation in the slope
of the $K^\pm \to \pi^+ \pi^- \pi^\pm$ Dalitz plot. Theoretical predictions
for CP violation lie between $10^{-4}$ and $10^{-6}$ for the difference 
of the slope parameter $g$ between $K^+$ and $K^-$ Dalitz plot. The sensitivity 
of NA48/2 is expected to be better than $10^{-4}$.

Further goals of the NA48/2 experiment are the absolute measurement of Br$(\kppienu)$
to determine the CKM matrix element $V_{us}$, a precise measurement of
$\kppipie$ $(K_{e4})$ decays, and the investigation of various rare $K^+$ decays.


\end{document}